# ReAcTable: Enhancing ReAct for Table Question Answering

Yunjia Zhang
University of Wisconsin-Madison
yunjia@cs.wisc.edu

Jordan Henkel
Microsoft
jordan.henkel@microsoft.com

Avrilia Floratou
Microsoft
avflor@microsoft.com

Joyce Cahoon
Microsoft
jcahoon@microsoft.com

Shaleen Deep
Microsoft
shaleen.deep@microsoft.com

Jignesh M. Patel*
Carnegie Mellon University
jignesh@cmu.edu

## ABSTRACT

Table Question Answering (TQA) presents a substantial challenge at the intersection of natural language processing and data analytics. This task involves answering natural language (NL) questions on top of tabular data, demanding proficiency in logical reasoning, understanding of data semantics, and fundamental analytical capabilities. Due to its significance, a substantial volume of research has been dedicated to exploring a wide range of strategies aimed at tackling this challenge including approaches that leverage Large Language Models (LLMs) through in-context learning or Chain-of-Thought (CoT) prompting as well as approaches that train and fine-tune custom models.

Nonetheless, a conspicuous gap exists in the research landscape, where there is limited exploration of how innovative foundational research, which integrates incremental reasoning with external tools in the context of LLMs, as exemplified by the ReAct paradigm, could potentially bring advantages to the TQA task. In this paper, we aim to fill this gap, by introducing ReAcTable (**ReAc**t for **Table** Question Answering tasks), a framework inspired by the ReAct paradigm that is carefully enhanced to address the challenges uniquely appearing in TQA tasks such as interpreting complex data semantics, dealing with errors generated by inconsistent data and generating intricate data transformations. ReAcTable relies on external tools such as SQL and Python code executors, to progressively enhance the data by generating intermediate data representations, ultimately transforming it into a more accessible format for answering the user's questions with greater ease. Through extensive empirical evaluations using three popular TQA benchmarks, we demonstrate that ReAcTable achieves remarkable performance even when compared to fine-tuned approaches. In particular, it outperforms the best prior result on the WikiTQ benchmark, achieving an accuracy of 68.0% without requiring training a new model or fine-tuning.

**PVLDB Artifact Availability:**
The source code, data, and/or other artifacts have been made available at https://github.com/yunjiazhang/ReAcTable.git.

## 1 INTRODUCTION

Table question answering (TQA) [16] is a subfield of natural language processing (NLP) and information retrieval that focuses on answering natural language (NL) questions over tabular data such as Wikipedia tables, spreadsheets or relational tables. It constitutes a complex task that demands a fusion of contextual understanding, logical reasoning and analytical skills. TQA allows users without expertise in querying languages and data analytics to interact with their data using plain language and gain valuable insights. It is a vital tool that can enhance data accessibility, usability, and decision support across various domains, ultimately leading to more efficient and informed decision-making processes.

Recognizing its significance, extensive research efforts have been dedicated to devising effective strategies for TQA. These strategies can be broadly classified into two categories. In the first category, approaches such as Tapas [12], Tapex [23], Tacube [57], and OmniTab [15] involve the training or fine-tuning of specialized models tailored for the task. The second category capitalizes on recent advancements in Large Language Models (LLMs). Within this category, works like [5, 26, 50] harness LLMs to generate code capable of manipulating tabular data.

The emergence of Chain-of-Thought (CoT) prompting, which encourages a model to engage in step-by-step reasoning, has brought about a significant transformation in the utilization of Large Language Models (LLMs) for intricate multi-step tasks. Expanding the CoT ideas, the ReAct paradigm [49] has been introduced, enabling interactions between the model and external tools in an interleaved manner. This allows for greater synergy between reasoning and acting and facilitates real-time guidance and corrections during task execution. These innovative strategies aim to address the limitations of traditional few-shot prompting methods [2]. Despite the promising results demonstrated by combining reasoning with external tools, to the best of our knowledge, the ReAct paradigm has not yet been applied to the TQA task.

This paper bridges this gap by investigating how the principles behind the ReAct framework, i.e. CoT and availability of external tools, can be applied to the TQA task. Beyond the anticipated difficulty of accurately comprehending the user's natural language query, the TQA task poses a series of distinct challenges, including: (i) interpreting potentially intricate data semantics, (ii) the presence of noisy or inconsistent data, and (iii) the necessity for complex data transformations to derive the correct response. Let us take the example table in Figure 1 drawn from the WikiTQ [29] dataset. In this example, the user wants to know *"Which country had the most cyclists finish in the top-10?"*. As we see in the corresponding table, there is no column containing explicitly information about countries. Instead, the Cyclist column contains values that encompass both the cyclist's name and an accompanying abbreviation in parentheses signifying their affiliated country, constituting a column with data values that encapsulate complex semantics, all condensed into a single string entry. For a developer to answer this question, they would have to first understand where the country information is represented in the data, come up with a transformation to extract it from Cyclist column and then write a query/program to filter the data by rank and then group them by the corresponding country by applying the transformation on each row of the table. This process is quite challenging to automate for arbitrary tables and columns especially as the complexity of the data and/or NL questions increases.

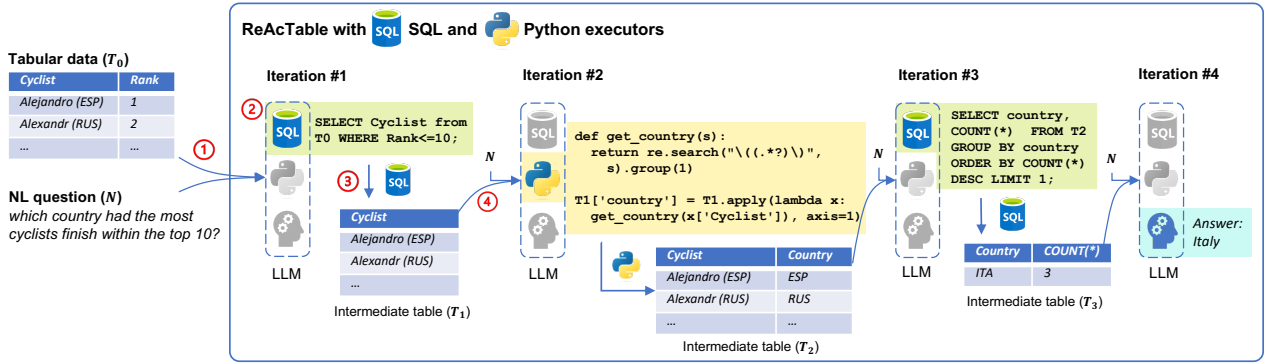

Figure 1: Overview of the ReAcTable framework with SQL and Python code executors.

To address the aforementioned challenges, we introduce ReAcTable (**ReAc**t for **Table** Question Answering tasks), a novel TQA framework inspired by the ReAct framework that combines step-by-step reasoning, code execution through external tools, intermediate table generation and majority voting. ReAcTable dissects intricate reasoning tasks into smaller, more manageable steps. These steps iteratively enhance the data by generating intermediate data representations, ultimately transforming it into a more accessible format for answering the user's questions with greater ease.

Figure 1 presents an overview of the ReAcTable framework. The framework employs two external tools: one for executing SQL queries and another for running Python code. Although the framework is adaptable to a range of code execution tools capable of manipulating tabular data, we have selected these two, as SQL serves as the standard language for querying structured data, and Python stands as the predominant choice among data scientists for data cleaning and transformation tasks [31]. The input for the ReAcTable framework comprises two key elements: (i) a tabular data set ($T_0$) and (ii) a natural language question ($N$) regarding $T_0$. ReAcTable automatically forms a prompt that is sent to a LLM (see Section 3.2 for details on the prompt). The LLM is equipped with three possible actions: (i) generating SQL code, (ii) generating Python code, or (iii) directly providing an answer. If the LLM opts to generate code, ReAcTable automatically uses the appropriate code executor to process the code and generate an intermediate table (e.g., $T_1$ in Figure 1). The process is repeated until the LLM produces a direct answer to the initial question (more details can be found in Section 3). It is crucial to note that this iterative process progressively refines the data, generating increasingly reliable context in the form of intermediate tables for subsequent reasoning iterations. As an example, by the end of the second iteration, the intermediate table contains a distinct Country column extracted from the original data, that allows the next iteration to easily group the data by country. Another intriguing observation is that the LLM employed SQL code for querying tabular data while opting for the Python executor when handling string manipulation tasks, a pattern that closely aligns with human preferences. Finally, as we discuss in Section 3.4, ReAcTable additionally leverages majority voting to improve the predictive quality of LLMs and also handle errors/code execution exceptions that might result from noisy/inconsistent data values.

The key contributions of the paper are:

- We introduce a novel framework, namely ReAcTable, for TQA tasks inspired by prior work on CoT and leveraging external tools (ReAct). ReAcTable utilizes large language models (LLMs) to break the problem into multiple steps, and generate logical operations in the form of code for processing tabular data when needed. This generated code is subsequently evaluated by external code executors, and the resulting intermediate table is fed back into the LLMs to facilitate subsequent reasoning steps. Additionally, ReAcTable leverages majority voting to improve the overall accuracy (see Section 3.4).

- We perform a thorough empirical evaluation of the ReAcTable framework using three popular TQA benchmarks (WikiTQ [29], TabFact [4], and FetaQA [25]) and compare it with various state-of-the-art approaches (LLM and non-LLM-based). We find that ReAcTable achieves remarkable performance across all the data sets. In particular, the framework achieves a test accuracy of 68.0% on WikiTQ [29], the most commonly used benchmark for TQA tasks, outperforming the state-of-the-art approaches by 2.1%, even when comparing with approaches that require fine-tuning.

- We delve deep into the behavior of ReAcTable and analyzing the contributions of each component (intermediate tables, availability of code executors, iterative processing) using an ablation study. We empirically demonstrate that the most significant improvement comes from the iterative generation of intermediate tables by the two code executors which aligns with our intuition that progressive refinement of the data can improve the LLMs predictive capabilities. We also carefully examine the effects of using different LLMs and majority voting mechanisms. More detailed results and analysis are shown in Section 4.3.

## 2 BACKGROUND

In the context of question-answering tasks, LLMs have become pivotal due to their ability to generate coherent and contextually relevant responses to the given questions. Despite the advancements in LLMs, "out-of-the-box" LLMs still struggle with complex questions[5, 46, 49]. To address these limitations, various individual methodologies have been proposed to enhance the traditional LLM approach. For example, ensemble methods, such as majority voting, have been employed to facilitate the exploration of multiple responses. Additionally, frameworks such as Chain-of-Thought (CoT) and ReAct have been introduced to improve the quality of LLM responses by altering the way we prompt the LLM. Since ReAcTableis built upon this foundational work, we provide related background and terminology in this section.

### 2.1 Large Language Models

Large Language Models (LLMs) represent a breakthrough in natural language processing, transforming the landscape of human-computer interaction and information processing. These models, often based on transformer architectures [44], such as GPT-3 [2], Codex [3], T5 [35], and Code Llama [37], are pre-trained on vast amounts of text data. By capturing context, these LLMs perform well in generating coherent and contextually relevant text, exhibiting capabilities ranging from text completion and language translation to question answering and code generation. The underlying mechanism involves attention mechanisms that allow models to weigh the significance of different words within a given context[44]. LLMs have demonstrated remarkable performance across numerous language tasks, including machine translation [2, 58], question answering [2, 36], etc.

### 2.2 Majority Voting Mechanisms

LLMs may produce responses that are uncertain or influenced by biases. Majority voting mechanisms in LLMs serve as a widely adopted approach to improve the response quality and mitigate potential biases. With majority voting mechanisms, multiple responses are sampled from the LLM's output distribution and one of the responses is selected based on some criteria. One common majority voting method is simple majority voting, which simply selects the response that occurs most frequently among the multiple generated outputs. Other criteria can be applied in majority voting methods, tailored to specific applications. For example, in code generation tasks, majority voting methods can also consider the execution results of generated code as a basis for selection [18, 26]. The choice of voting method depends on the context and requirements of the application.

### 2.3 Chain-of-Thought and ReAct Framework

In addition to simply prompting the LLMs, multiple prompting paradigms are designed to enhance the quality of reasoning and responses generated by these models. One specific example is the Chain-of-Thought (CoT) prompting [46], which goes beyond traditional prompting and introduces a structured approach to guide the model's reasoning process. In the CoT paradigm, the reasoning process is organized into multiple intermediate steps, allowing the model to solve one simpler subproblem at a time and progressively build a coherent response. This structured approach helps LLMs tackle complex tasks effectively.

The ReAct framework [49] has expanded upon the foundational ideas of the CoT paradigm, introducing the concept that interactions with external components can substantially enhance the capabilities of LLMs. By enabling LLMs to engage with external components, ReAct broadens the scope of tasks and applications that these models can handle, making them more versatile and adaptable. A similar idea has also been introduced into modern LLMs, including GPT-4 [27], enabling them to provide users with better results via external plugins.

### 2.4 Table Question Answering

Table Question Answering (TQA) is a task that resides at the intersection of natural language processing and data analytics. In this paper, our focus is on solving the single-table TQA problem. Specifically, when presented with a natural language question $N$ about a relational table $T_0$, our objective is to provide the correct answer to $N$. The output answer may take the form of a tuple list or a sentence in natural language. For instance, the WikiTQ [29] benchmark presents answers in tuple lists (e.g., "2001|2002|2003"), while the FeTaQA [25] benchmarks uses natural language as the answer format (e.g., "Harvey beat Royds by 1,463 votes"). We assume that both schema-level information and data content of the tabular data $T_0$ are available to answer the given question $N$.

## 3 ReAcTable: ReAct FOR TQA TASKS

ReAcTable (**ReAc**t for **Table** Question Answering tasks) is an instantiation of the ReAct framework [49], designed to tackle complex, multi-step Table Question Answering (TQA) problems (as defined in Section 2.4). By incorporating concepts from ReAct, along with majority voting mechanisms and specialized code executors, ReAcTable is able to break down complex TQA problems into smaller, simpler sequenced tasks. For each of these sequenced tasks, it employs generated code to manipulate the target table. Thus, ReAcTable significantly boosts the performance of pre-trained Large Language Models (LLMs) without additional fine-tuning.

### 3.1 Overview

Figure 1 provides an overview of the ReAcTable framework. The inputs to ReAcTable consist of (i) a relational data table $T_0$, and (ii) a natural language question $N$. The ultimate goal of ReAcTable is to produce the correct answer to the given question.

ReAcTable iteratively separates the complex TQA task into smaller tasks. At each iteration, the tabular data ($T_0$) and the natural language question ($N$) are first input into the LLM (① in Figure 1). The LLM can then perform one of three distinct operations (② in Figure 1): (i) generating a SQL query, (ii) generating Python code, or (iii) directly answering the question.

The ReAcTable framework is designed to be adaptable, allowing for the integration of other code executors in addition to SQL and Python. In our work, we use these two specific executors as they are commonly used by data scientists to manipulate tabular data. If the LLM generates SQL or Python code, ReAcTable activates the corresponding executor. The execution results are in an intermediate data table (③ in Figure 1). This table is derived from the initial tabular data and is tailored to address the question more directly. The intermediate table, along with the original natural language question ($N$), is then fed back into the LLM for subsequent iterations (④ in Figure 1). The iterative process continues until the LLM provides a direct answer to the question (instead of generating code for an executor). To enable the LLM to "learn" how to answer the question, we employ the in-context learning paradigm [2], using (static) few-shot examples in our prompts. The formulation of these prompts is elaborated in Section 3.2.

For the example question illustrated in Figure 1 ("*which country had the most cyclists finish within the top 10*"), ReAcTable employs four iterations to generate the answer. This TQA example is sourced from the WikiTQ data set [29]. In the initial iteration, ReAcTable generates SQL code to select the cyclists who finished the race within the top 10 ranks. This SQL code is further executed, producing an intermediate table $T_1$. Subsequently, using the intermediate table $T_1$ as input, ReAcTable's LLM generates Python code to extract the country code from the Cyclist column for each row. The python code is executed producing table $T_2$. For the third iteration, ReAcTable generates SQL code to count how many times each country appears in the intermediate table $T_2$, producing table $T_3$. Finally, in the last iteration, ReAcTable leverages the LLM to generate a natural language answer based on the country code found in $T_3$. Through these executors, ReAcTable improves the problem-solving capabilities of the LLM by effectively using intermediate tables ($T_1$, $T_2$, and $T_3$).

### 3.2 Prompting Large Language Models

***Prompt template.*** ReAcTable uses a prompt template that is instantiated on every iteration based on the output of the LLM at the previous iteration. The very first prompt contains the original table $T_0$ (the schema, values and data format) and the user's NL question. It also instructs the LLM to think step-by-step (CoT) and utilize the two available code executors as needed. The top part of Figure 2 reflects this prompt. Additionally, it contains few-shot demonstrations (not included in the figure because of space requirements) whose format we describe at the end of this section.

After each iteration, ReAcTable would have generated either the answer to the question or an intermediate table. In case an intermediate table is generated, it is leveraged to instantiate the prompt template at the follow-up iteration. Specifically, in the ongoing iteration, ReAcTable enhances the prompt used in the previous round

```
ReAcTable Prompt Example
The database table T0 is shown as follows:
[HEAD]:Rank|Cyclist|Team|Time|Uci_protour_points
---
[ROW] 1: 1|Alejandro Valverde (ESP)|Caisse d'Epargne|5h 29' 10\",40|NULL
[ROW] 2: 2|Alexandr Kolobnev (RUS)|Team CSC Saxo Bank|s.t.|30.0
...
[ROW] 10: 10|David Moncoutié (FRA)|Cofidis|+ 2\",1"|NULL

Answer the following question based on the data above: "which country had
the most cyclists finish within the top 10?". Generate SQL or Python code
step-by-step given the question and table to answer the question correctly.

ReAcTable: SQL: ```SELECT Cyclist FROM T0 WHERE rank<=10;```.

Intermediate table (T1):
[HEAD]: Cyclist
---
[ROW] 1: Alejandro Valverde (ESP)
[ROW] 2: Alexandr Kolobnev (RUS)
...
[ROW] 10: David Moncoutié (FRA)

ReAcTable: Python: ```
def get_country(s):
    return re.search("\((.*?)\)", s).group(1)
T1['Country'] = T1.apply(lambda x: get_country(x['Cyclist']), axis=1)```.

Intermediate table (T2):
[HEAD]: Cyclist|Country
---
[ROW] 1: Alejandro Valverde (ESP)|ESP
[ROW] 2: Alexandr Kolobnev (RUS)|RUS
...
[ROW] 10: David Moncoutié (FRA)|FRA

ReAcTable: SQL: ```SELECT Country, COUNT(*) FROM T2 GROUP BY Country ORDER
BY COUNT(*) DESC LIMIT 1;```.

Intermediate table (T3):
[HEAD]: Country|COUNT(*)
---
[ROW] 1: ITA|3                                          LLM generated:
                                                        Code executor generated:
ReAcTable: Answer: ```Italy```.
```

**Figure 2: ReAcTable prompt after the $4^{th}$ iteration for our running example.**

by incorporating the SQL or Python code generated in the preceding step, alongside the intermediate table formed by executing this code. This updated prompt functions as the input for the present iteration, guaranteeing that the LLM possesses a comprehensive view of the related context at the current stage.

Figure 2 displays the prompt created by ReAcTable at the end of the fourth iteration shown in Figure 1. As shown in the figure this prompt contains all the information used in the very first prompt (top part of the figure) along with the LLM output (green) and intermediate table (yellow) produced at each subsequent iteration.

Note that ReAcTable is not restricted to using the "SQL-Python-SQL-Answer" pattern shown in Figure 1. It can opt for any combination of executors and the prompt template will be initiated accordingly. For straightforward questions with intuitive answers, ReAcTable may not generate any intermediate code, while for complex questions, ReAcTable may employ five or more iterations to break down the complexity into smaller sub-questions. This flexibility allows ReAcTable to adapt to questions of varying complexity.

***Few-shot demonstrations.*** In-context learning through few-shot demonstrations is a common technique for calibrating LLMs to specific tasks [2]. By utilizing a small set of examples that closely represent the problem structure, the LLM can be adapted to the task at hand at *inference time*. In ReAcTable, we also use few-shot demonstrations to guide LLMs. These demonstrations are inserted at the beginning of the prompt template along with the original table and NL question. These examples are ommitted from Figure 2 in the interest of space but follow the format shown in the figure.

## 3.3 Interacting with External Code Executors

***Code execution.*** Once the LLM used by ReAcTable generates code, the system initializes corresponding code executors to execute the code using the provided table in the current context. In the example shown in Figure 1, the table is represented as a SQLite [13] table when running SQL queries, while a Python Pandas DataFrame [24] is used to run Python code.

***Handling exceptions.*** Given that the code is generated by the LLM, there remains a possibility of encountering exceptions during code execution. Effectively managing such exceptions is crucial for ensuring the "last-mile" performance of ReAcTable. Common exceptions are addressed as follows:

- **SQL exceptions**: In the SQL running context, a common error is that the SQL query requires a column that does not exist in the given table. However, such a column may exist in previous intermediate tables or the original table. To address this, we introduce a retry mechanism that allows SQL queries on previous intermediate tables in reverse order. If the query successfully executes on a table, the result table is used as the next intermediate table of ReAcTable. In addition, to further reduce the execution error caused by column names, the column names are normalized by removing spaces, leading numbers, and special characters.

- **Python "module not found" exception**: In the Python running environment, we import common packages like regular expressions (re) and date time (datetime), etc. If a new module is used in the generated Python code, we install the package at runtime and rerun the Python code.

- **Other exceptions**: For all other exceptions, we "force" the LLM to generate an answer by appending the leading word "Answer:" at the end of the prompt.

## 3.4 Voting Mechanisms

Majority voting mechanism is frequently used to improve the predictive quality of LLMs, as it facilitates the exploration of various responses and more effectively addresses ambiguity [18]. Since the core of ReAcTable lies an LLM, it can integrate with different voting mechanisms to increase the accuracy of the prediction. In this section, we describe three different voting strategies with which the ReAcTable framework can effectively operate: (i) simple majority voting, (ii) tree-exploration voting, and (iii) execution-based voting. Figure 3 shows an overview of these voting mechanisms.

### 3.4.1 Simple majority voting.

Figure 3a shows the simple majority voting mechanism used in ReAcTable. We use chains to represent the problem-solving steps in ReAcTable, with nodes representing the programs or answers predicted by the LLM. In simple majority voting, we apply a high-temperature setting to the LLM [41] and perform ReAcTable's step-by-step solving iterations for a total of $n$ times. As a result, we obtain $n$ predicted answers. The majority answer among these predictions is selected as ReAcTable's final prediction. A summary of the simple majority voting mechanism is shown in Algorithm 1.

---

**Algorithm 1** ReAcTable with simple majority voting

**Input:** Table $T_0$, Question $N$, Temperature $t$, Sample time $n$
**Output:** Predicted answer $P$
1. $answers = []$
2. for $n$ times:
3.    $tabs = [T_0]$
4.    while true:
5.       $prompt = preparePrompt(tabs, N_0)$
6.       $pred = LLM(prompt, t)$ # get LLM prediction
7.       if $pred$ is code:
8.          $T' = Execute(pred, tabs)$ # execute the code
9.          $tabs.append(T')$ # log the intermediate table
10.      else:
11.         $answers.append(pred)$
12.         break
13. return $getMajority(answers)$

---

**Algorithm 2** ReAcTable with tree-exploration voting

**Input:** Table $T_0$, Question $N$, Temperature $t$, Sample time $n$
**Output:** Predicted answer $P$
1. $tabs = [T_0]$, $answers = []$
2. $treeBranches = initQueue()$ # a queue to store branches
3. $treeBranches.append(tabs)$
4. while $treeBranches$ is not empty:
5.    $tabs' = treeBranches.popLeft()$
6.    $prompt = preparePrompt(tabs', N_0)$
7.    $all\_preds = LLM(prompt, t, n)$
8.    for $pred$ in $all\_preds$:
9.       if $pred$ is code:
10.         $T' = Execute(pred, tabs')$
11.         $treeBranches.append(tabs' + [T'])$
12.      else:
13.         $answers.append(pred)$
14. return $getMajority(answers)$

---

### 3.4.2 Tree-exploration voting.

Figure 3b demonstrates the tree-exploration voting mechanism. The idea behind tree-exploration voting is to enable LLMs to explore multiple intermediate steps before arriving at the final answer. Unlike simple majority voting, which repeats the entire chain multiple times, tree-exploration voting allows the LLM to sample $n$ times at each prediction, resulting in a fanout of $n$ in the reasoning tree. In this voting mechanism, ReAcTable traverses all branches of the tree until each branch reaches an answer. Finally, ReAcTable selects the majority of the answers as the final prediction. The algorithm for tree-exploration voting is shown in Algorithm 2.

### 3.4.3 Execution-based voting.

In the two voting methods above, although they recognize the data context from earlier stages (as stated in the prompt), it does not consider the output of the code (intermediary result table) when making decisions on which prediction to take. To utilize the data context for selecting the next step, we introduce an execution-based voting mechanism [18, 26].

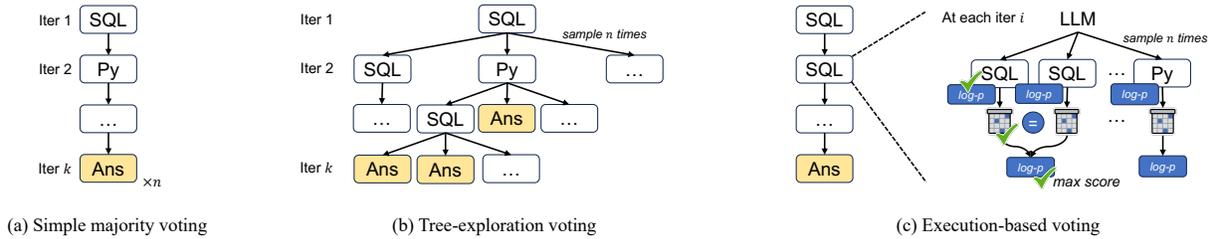

Figure 3: Overview of voting mechanisms

**Algorithm 3** ReAcTable with execution-based voting
**Input:** Table $T_0$, Question $N$, Temperature $t$, Sample time $n$
**Output:** Predicted answer $P$
1. $tabs = [T_0]$
2. while true:
3.     $prompt = preparePrompt(tabs, N_0)$
4.     $all\_preds = LLM(prompt, t, n)$
5.     $resultLog = initResultLog()$ # initialize the result log
6.     for $pred, log\_prob$ in $all\_preds$:
7.       if $pred$ is code:
8.          $resultLog.update(Execute(pred, tabs), log\_prob)$
9.       else:
10.          $resultLog.update(pred, log\_prob)$
11.     # get prediction with max score
12.     $pred = resultLog.get\_prediction()$
13.     if $pred$ is code:
14.       $T' = Execute(pred, tabs)$ # execute the code
15.       $tabs.append(T')$ # log the intermediate table
16.     else:
17.       return $pred$

In execution-based voting, ReAcTable allows the LLM to sample $n$ predictions at each reasoning step. Instead of exhaustively exploring the tree like tree-exploration voting, execution-based voting selects only one of the predictions as the intermediate step. For each of the $n$ predictions, if the prediction is a program, ReAcTable executes the program and retrieves the resulting table. If equivalent tables are produced, the log probabilities are merged by selecting the maximum log probability (log-p in Figure 3c). Finally, ReAcTable selects the code or answer with the highest score as the next step in the reasoning chain. This decision-making process is repeated for each step in the question-answering procedure. Algorithm 3 shows a step-by-step explanation of execution-based voting.

### 3.5 Discussion

***Comparing ReAcTable with CoT and ReAct.*** ReAcTable can be viewed as an advanced and specialized version of the CoT and ReAct paradigms, specifically tailored to address the unique challenges and requirements of TQA tasks. The primary distinction between these frameworks and ReAcTable lies in ReAcTable's utilization of code executors, which are also essential tools for data scientists, to generate intermediate tables. When working with these intermediate tables which are progressively refined, the LLM produces more coherent and semantically correct code to solve the user question in a step-by-step fashion. Furthermore, ReAcTable incorporates specific design elements crucial for solving TQA problems, such as handling execution exceptions and integrating with majority voting methods. ReAcTable demonstrates that the CoT paradigm is also performance-critical in the TQA scenario, providing data scientists with an easily implementable and high-performing framework.

***Comparing different majority voting methods.*** In ReAcTable, simple majority voting and tree-exploration majority voting can explore a broader range of possible solution paths. By generating multiple code segments and finally selecting the answers that occur the most frequently, they enable the exploration of various solutions to the given question. This naturally aligns with the observation that there could be multiple solutions that lead to the same correct answer. In contrast, execution-based majority voting prioritizes the selection of code segments that are more likely to produce semantically correct results when executed. Instead of exploring the diverse solutions to the same question, this approach emphasizes the practicality and correctness of the generated code.

The selection of the majority voting methods in ReAcTable depends on multiple factors, including the complexity of the TQA task and the capabilities of the underlying LLM. In this paper, we empirically compare these majority voting methods in Section 4.2 and demonstrate that all majority voting methods result in improved performance compared to ReAcTable without majority voting. Additionally, we demonstrate that the selection of the optimal voting method is also a non-trivial task as it depends on the capabilities of the underlying LLM (see 4.4). We discuss this aspect in our directions for future work.

## 4 EXPERIMENTAL EVALUATION

In this section, we present the experimental evaluation of ReAcTable. Our findings show that ReAcTable surpasses state-of-the-art approaches across multiple commonly used TQA datasets. To dive deeper into understanding why ReAcTable achieves better performance, we conduct comprehensive ablation studies. Furthermore, we showcase the versatility of ReAcTable by demonstrating its compatibility with various GPT-series language models, highlighting its adaptability to different model architectures and characteristics.

### 4.1 Experimental Setup

***Benchmarks.*** We adopt three commonly used benchmarks – WikiTQ [29], TabFact [4], and FeTaQA [25] – to evaluate ReAcTable against state-of-the-art baseline approaches. For all three data sets, we use the training sets to create static few-shot examples for ReAcTable and use the same sets of few-shot examples throughout the paper.
- **WikiTQ:** WikiTableQuestions (WikiTQ) is a data set designed to facilitate research in the field of question answering over structured tabular data [29]. The WikiTQ data set comprises 14,149 question-answer pairs in the training set and 4,344 in the test set. The answers in the WikiTQ data set can take three forms: (i) natural language answers derived from a single tuple, (ii) lists of values extracted from multiple tuples, or (iii) analytical answers that do not exist in the tables.
- **TabFact:** Table-Fact-Checking (TabFact) data set provides a diverse collection of tables sourced from various domains, accompanied by a set of fact-checking queries [4]. The answers to the given queries in TabFact are binary ("yes" or "no"), indicating whether the given queries state facts or not based on the tabular data. To reduce the experimental cost without losing generality, we chose to use the small test set provided [4] to evaluate the performance of ReAcTable and all baseline approaches [5]. The small test set contains 1,998 question-answer pairs.
- **FeTaQA:** Free-form Table Question Answering (FeTaQA) is a free-form question-answering data set built upon Wikipedia and presents a different TQA scenario – answering the question with natural languages [25]. FeTaQA contains 7,326 question-answer pairs in the training set and 2,006 in the test set.

**Baselines.** There are two categories of baseline approaches: (i) approaches that require training, and (ii) approaches that do not require training. Because they differ in their training data requirements, we report these two categories separately. For the approaches that require training, we include Tapex [23], TaCube [57], OmniTab [15], TaPas [12], SaMoE [54], PASTA [10], Lever [26], and the T5 series of models (T5-Small, T5-Base, and T5-Large) [34]. Among these approaches, TaPex, OmniTab, TaPas, SaMoE, and PASTA pretrain their own models with custom training data, while TaCube and T5 are fine-tuned on existing pre-trained large language models. In terms of approaches that do not require training, we report Binder [5] and Dater [50] due to their strong performance. Because the best-performing baseline varies across different benchmarks, we report the best-performing baselines for each benchmark. It is also worth noting that these LLM-based baseline approaches also incorporate various majority voting methods as described in [5, 50]. For all the baseline approaches, we report the best results observed.

**ReAcTable configurations.** In our experiments, we report the performance of ReAcTable with and without the three majority voting methods. We denote ReAcTable without majority voting as *ReAcTable*, ReAcTable with simple majority voting as *ReAcTable with s-vote*, ReAcTable with tree-exploration majority voting as *ReAcTable with t-vote*, and ReAcTable with execution-based majority voting as *ReAcTable with e-vote*. Regarding the temperature parameter in LLMs [41], we employ two settings: We set the LLM temperature to zero for *ReAcTable*, while for all experiments utilizing majority voting, we consistently set the temperature to 0.6, which is the common setting of previous works [18, 26]. Additionally, for the code execution environment, we use SQL and Python executors in ReAcTable, unless otherwise specified (Section 4.3.3).

Table 1: Performance of ReAcTable on WikiTQ data set.

| Methods | Accuracy |
| --- | --- |
| *Approaches require training* | |
| Tapex | 57.5% |
| TaCube | 60.8% |
| OmniTab | 62.8% |
| Lever | **62.9%** |
| *Approaches without training* | |
| Binder | 61.9% |
| Dater | 65.9% |
| ReAcTable | 65.8% |
| with s-vote | **68.0%** |
| with t-vote | 66.4% |
| with e-vote | 67.2% |

Table 2: Performance of ReAcTable on TabFact data set.

| Methods | Accuracy |
| --- | --- |
| *Approaches require training* | |
| TaPas | 83.9% |
| Tapex | 86.7% |
| SaMoE | 86.7% |
| PASTA | **90.8%** |
| *Approaches without training* | |
| Binder | 85.1% |
| Dater | 85.6% |
| ReAcTable | 83.1% |
| with s-vote | **86.1%** |
| with t-vote | 84.2% |
| with e-vote | 84.9% |

For the underlying LLM of ReAcTable, we use Codex [3, 42] as the default LLM. To evaluate the versatility of ReAcTable, we also evaluate ReAcTable with other GPT-series LLMs in Section 4.4.

**Metrics.** We mainly use accuracy to compare the response quality of ReAcTable with the baseline approaches. Since the table question answering task can give multiple tuples as output answers, we use set-based comparison to evaluate the output answer against the given gold answer. To evaluate the quality of the WikiTQ data set, we use the official Python-based WikiTQ evaluator [47]. For the TabFact data set, since the answer is binary ("yes" or "no"), we simply use string matching. As for FeTaQA, since the gold answers are free-form natural language-based answers, we use the commonly adopted ROUGE-N and ROUGE-L metrics [22] to evaluate the "similarity" of predicted answers to the gold answers.

### 4.2 Performance of ReAcTable

In this section, we compare the performance of ReAcTable with state-of-the-art approaches using three commonly used TQA data sets: WikiTQ, TabFact, and FetaQA.

Table 3: Performance of ReAcTable on FeTaQA data set.

| Methods | ROUGE-1 | ROUGE-2 | ROUGE-L |
| --- | --- | --- | --- |
| *Approaches require training* | | | |
| T5-Small | 0.55 | 0.33 | 0.47 |
| T5-Base | 0.61 | 0.39 | 0.53 |
| T5-Large | 0.63 | 0.41 | 0.53 |
| *Approaches without training* | | | |
| Dater | 0.66 | 0.45 | 0.56 |
| ReAcTable | **0.71** | **0.46** | **0.61** |

Table 4: ReAcTable vs. Codex with simple chain-of-thought on WikiTQ data set.

| Methods | Accuracy |
| --- | --- |
| Codex-CoT | 49.4% |
| with s-vote | 47.7% |
| ReAcTable | 65.8% |
| with s-vote | **68.0%** |

**Results.** Table 1, 2, and 3 show the performance of ReAcTable on WikiTQ, TabFact, and FetaQA, respectively.

As shown in Table 1, *ReAcTable with s-vote* achieves an accuracy of 68.0%, outperforming all baseline approaches (including both fine-tuned approaches and approaches without fine-tuning). It is worth noting that the results of the baseline approaches also incorporate various majority voting methods [5, 26]. For *ReAcTable* (without majority voting), it still maintains an accuracy of 65.8%. Among the three majority voting methods, *ReAcTable with s-vote* performs the best (68.0%), while *ReAcTable with t-vote* exhibits a relatively lower accuracy (66.4%). However, all three majority voting methods (*ReAcTable with s-vote, t-vote, and e-vote*) improve the performance of the original *ReAcTable*.

Turning our attention to the TabFact data set, as shown in Table 2, we also observe that *ReAcTable with s-vote* outperforms all state-of-the-art approaches without fine-tuning. Regarding the approaches with fine-tuning, *ReAcTable with s-vote* is still 4.7% lower than the best-performing baseline.

Regarding FeTaQA, as shown in Table 3, given that the gold answers to the questions are free-form natural language sentences, we use the commonly adopted ROUGE-1, ROUGE-2, and ROUGE-L as evaluation metrics, where higher values represent better results. *ReAcTable* achieves the highest ROUGE score compared to any other reported baseline.

**Takeaways.** As shown in the above results, ReAcTable consistently outperforms the state-of-the-art approaches on commonly used benchmarks. It is also worth noting that, in contrast to fine-tuning-based approaches, ReAcTable does not require any training steps, making ReAcTable easier to implement and deploy.

Regarding various voting mechanisms, we observe that ReAcTable with simple majority voting outperforms the other two voting methods. Beyond this experiment, we also observe that these majority voting methods may perform differently when using different LLMs (see Section 4.4). Since selecting the best voting method is also a non-trivial research topic, we list this as one of the future directions.

## 4.3 Analyzing ReAcTable

In this section, we aim to analyze ReAcTable to address a fundamental question: *Why does ReAcTable outperform the state-of-the-art approaches*. Considering that the major difference between ReAcTable and the traditional CoT paradigm is whether using intermediate tables to facilitate the subsequent reasoning steps, in this section, we first analyze how the intermediate tables affect the performance of ReAcTable. Then, we investigate whether controlling the maximum number of iterations has an impact on the performance of ReAcTable. Finally, we analyze how the choice of different code executors (SQL and Python) affects the results.

### 4.3.1 Effect of intermediate tables.

To analyze how intermediate tables affect the performance of *ReAcTable*, we conducted an ablation study of *ReAcTable* by removing the intermediate tables and creating a new method named *Codex-CoT*. *Codex-CoT* simply generates a sequence of code over the tabular data with a single text completion iteration using the LLM (Codex). Subsequently, the generated code is executed to obtain the final answer to the question. As the only difference between *Codex-CoT* and *ReAcTable* lies in whether they use intermediate tables as the code generation context, a direct comparison between the two approaches enables us to understand the impact of intermediate results on end-to-end performance.

As shown in Section 4.2, simple majority voting (*ReAcTable with s-vote*) stands out as the best-performing configuration among various majority voting methods. Therefore, in this experiment, we exclusively focus on the simple majority voting setup, specifically comparing *Codex-CoT with s-vote* with *ReAcTable with s-vote*.

**Results.** Table 4 and Table 5 present the results of our ablation study on the WikiTQ and TabFact data sets, respectively.

From Table 4, we observe that *Codex-CoT* achieves an accuracy of only 49.4%, while *ReAcTable* performs notably better with an accuracy of 65.8%, which is a 16.4% improvement. Interestingly, when applying simple majority voting to *Codex-CoT*, the accuracy drops to 47.7%. One possible reason for this decline is that when the LLM is uncertain about the answer, using a high-temperature setup (0.6 in *Codex-CoT with s-vote*) can further increase uncertainty, leading to worse results compared to the low-temperature setup [18].

Regarding the TabFact data set, we make a similar observation from Table 5, where *Codex-CoT* exhibits a 12.0% lower accuracy compared to *ReAcTable*. Furthermore, even after applying simple majority voting, *Codex-CoT with s-vote* achieves an accuracy of 72.3%, which is still 13.8% behind the accuracy of *ReAcTable with s-vote*.

**Takeaways.** Our ablation study results demonstrate that the inclusion of intermediate tables significantly contributes to the performance enhancement of ReAcTable. Moreover, it is worth noting that majority voting mechanisms do not consistently yield better accuracy, especially when the LLM is uncertain about the answer.

Table 5: ReAcTable vs. Codex with simple chain-of-thought on TabFact data set.

| Methods | Accuracy |
| --- | --- |
| Codex-CoT | 71.1% |
| with s-vote | 72.3% |
| ReAcTable | 83.1% |
| with s-vote | **86.1%** |

Table 6: Accuracy breakdown of ReAcTable on WikiTQ data set. We show the accuracy when ReAcTable uses different numbers of iterations to solve the questions, i.e. the accuracy for each bar in Figure 4a.

| Iteration # used by ReAcTable | Accuracy |
| --- | --- |
| Iteration # = 1 (# of data points: 233) | 62.8% |
| Iteration # = 2 (# of data points: 3,426) | 72.3% |
| Iteration # = 3 (# of data points: 364) | 60.3% |
| Iteration # = 4 (# of data points: 264) | 59.3% |
| Iteration # = 5 (# of data points: 19) | 46.2% |

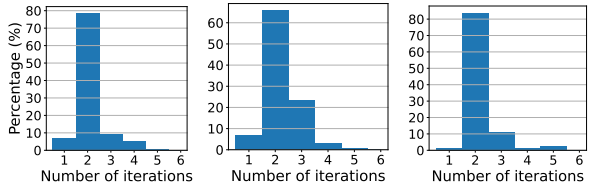

(a) WikiTQ  (b) TabFact  (c) FeTaQA

Figure 4: Distribution of the number of iterations.

Table 7: Performance of ReAcTable on WikiTQ data set.

| Methods | Accuracy |
| --- | --- |
| ReAcTable (*with s-vote*) | |
| *limit* = 1 | 49.2% |
| *limit* = 2 | 65.1% |
| *limit* = 3 | 67.3% |
| *unlimited* | **68.0%** |

### 4.3.2 Number of iterations.

ReAcTable uses multiple rounds of interactions with LLMs and code executors to handle complex question-answering tasks over tabular data. In ReAcTable, the number of iterations can affect the performance. On one hand, more iterations may enable ReAcTable to reason through questions. On the other hand, increased iterations may also indicate that the question is complex and hard to answer. To gain insights into how the number of iterations affects ReAcTable performance, we aim to answer two questions: 1) *How many iterations does ReAcTable utilize when we do not control the number of iterations*, and 2) *How does controlling the maximum iteration number impact the results*. In this experiment, we continue to evaluate ReAcTable with the three benchmarks: WikiTQ, TabFact, and FeTaQA.

***Number of iterations.*** We first analyze the number of iterations when we allow an *unlimited number of iterations* for ReAcTable. In this experiment, we use *ReAcTable with s-vote* on the three data sets – WikiTQ, TabFact, and FeTaQA – to perform the study. Figure 4 illustrates the distribution of interation numbers. As shown in Figure 4, across the three data sets, all questions are resolved within five iterations, with over 70% of the questions being answered within two iterations.

Furthermore, we break down the accuracy of *ReAcTable with s-vote* on the WikiTQ data set (shown in Figure 4a) to reveal the detailed accuracy corresponding to the different numbers of iterations chosen by ReAcTable (w.r.t. different bars in Figure 4a). As demonstrated in Table 6, ReAcTable achieves its highest performance when it opts for two iterations (72.3%). However, as the number of iterations increases, the performance of ReAcTable gradually declines. This observation suggests that questions that require more iteration steps might be inherently challenging for ReAcTable to handle, resulting in lower accuracy.

***Limiting the number of iterations.*** Next, we investigate whether imposing a limit on the number of iterations in ReAcTable impacts its performance. To establish a maximum iteration limit at $k$, we terminate the reasoning process at iteration $k-1$ if ReAcTable does not opt to directly answer the question. When reaching iteration $k$, we force ReAcTable to directly answer the question by appending the leading word "Answer" to the prompt.

Table 7 presents the results of ReAcTable when various maximum iteration limits are imposed. As the maximum iteration limit is raised, the accuracy of ReAcTable also rises. An interesting observation is that with a maximum limit set at two iterations, the accuracy reaches 65.1% (from 49.2%). Beyond this point, when the maximum limit exceeds two, the increase in accuracy becomes less pronounced. This observation aligns with the finding that a substantial portion of questions can be effectively answered within two iterations. In addition, from these results, imposing a maximum limit on the iteration number does not yield improvements in accuracy, as it could potentially restrict ReAcTable's capacity for reasoning over complex, multi-step questions.

***Takeaways.*** Based on the aforementioned results, the majority of questions in the three commonly used data sets can be answered in just two iterations. Moreover, for complex question-answering tasks where ReAcTable employs more than two iterations to generate an answer, restricting the number of iterations does not lead to improved accuracy. This shows the adaptability of ReAcTable to the complexity of questions and highlights the importance of allowing ReAcTable's flexibility in the iteration process when handling complex queries.

### 4.3.3 Effect of Using Different Code Executors.

In the default configuration of ReAcTable, two external executors are utilized: the SQL executor and the Python executor. The SQL code generated by ReAcTable primarily handles data selection, while the Python code generated by ReAcTable deals with data formatting. In this experiment, our objective is to assess the individual performance contributions of each executor.

Table 8: Performance of ReAcTable on WikiTQ data set with only the SQL query executor.

| Methods | Accuracy |
| --- | --- |
| ReAcTable | 65.8% |
| with s-vote | **68.0%** |
| with t-vote | 66.4% |
| with e-vote | 67.2% |
| ReAcTable (*with only the SQL executor*) | 62.5% |
| with s-vote | 64.5% |
| with t-vote | 64.1% |
| with e-vote | 63.6% |

To conduct these experiments, we modified the default *ReAcTable* configurations by removing the Python executor, resulting in a variant named *ReAcTable (with only the SQL executor)*. In this modified setup, *ReAcTable (with only the SQL executor)* exclusively employs SQL as the programming language for manipulating tabular data. Without the Python executor, ReAcTable may face challenges in reformatting the data, relying on the LLM to comprehend such challenges. It is worth noting that we only apply the ablation study to the Python executor since, in our setup, ReAcTable relies on the SQL executor to perform fundamental tabular data operations.

***Results.*** Table 8 and Table 5 present the results on the WikiTQ and TabFact data sets, respectively. When using *ReAcTable with only the SQL executor*, the accuracy on WikiTQ drops from 65.8% to 62.5%. After applying simple majority voting (*ReAcTable with s-vote*), removing the Python executor results in a decrease of 3.5%. For the WikiTQ test example question shown in Figure 1 and 2, although *ReAcTable with only the SQL executor* initially selects the same SQL code as the original *ReAcTable* in the first iteration, it chooses a direct answer, "Spain", in the second iteration, which is not shown to be correct based on the given table. Similarly, on TabFact, removing the Python executor leads to a significant drop in accuracy, up to 9.9%. This highlights the importance of the Python executor in enhancing the performance on these data sets.

***Takeaways.*** From our results, the inclusion of the Python executor indeed improves accuracy, particularly by allowing complex data reformatting via Python. It is also important to note that ReAcTable exhibits flexibility and can be adapted to work with various code executors beyond the default SQL and Python executors. We further discuss the choice of code executors in Section 5.2.

## 4.4 ReAcTable with Various Language Models

Currently, the landscape of LLMs is diverse and quickly evolving. Likewise, ReAcTable is versatile in its ability to improve the prediction quality of various LLMs. To illustrate the effectiveness of ReAcTable in different LLMs, we evaluate ReAcTable using various GPT series language models.

From the large number of models offered by OpenAI [28], we opted for two commonly used models in addition to the default *code-davinci-002* model: *text-davinci-003* and *gpt3.5-turbo* [43]. *text-davinci-003* is a recent and improved version of the LLM, tailored for tasks that require precise instructions. In contrast, *gpt3.5-turbo* stands as the most capable iteration of the GPT-3.5 model, optimized for cost-effective chat applications. In this experiment, we simply apply different LLMs for ReAcTable without any other changes. We still use WikiTQ and TabFact to perform the evaluation.

Table 9: Performance of ReAcTable on TabFact data set with only the SQL query executor.

| Methods | Accuracy |
| --- | --- |
| ReAcTable | 83.1% |
| with s-vote | **86.1%** |
| with t-vote | 84.2% |
| with e-vote | 84.9% |
| ReAcTable (*with only the SQL executor*) | 75.4% |
| with s-vote | 76.2% |
| with t-vote | 77.1% |
| with e-vote | 75.8% |

Table 10: Performance of ReAcTable with various GPT-series models on WikiTQ data set.

| Methods | Accuracy |
| --- | --- |
| ReAcTable (code-davinci-002) | 65.8% |
| with s-vote | 68.0% |
| with t-vote | 66.4% |
| with e-vote | 67.2% |
| ReAcTable (text-davinci-003) | 63.3% |
| with s-vote | 64.1% |
| with t-vote | 64.5% |
| with e-vote | 65.0% |
| ReAcTable (gpt3.5-turbo) | 52.4% |
| with s-vote | 51.8% |
| with t-vote | 52.5% |
| with e-vote | N.A.[1] |

Table 11: Performance of ReAcTable with various GPT-series models on TabFact data set.

| Methods | Accuracy |
| --- | --- |
| ReAcTable (code-davinci-002) | 83.1% |
| with s-vote | **86.1%** |
| with t-vote | 84.2% |
| with e-vote | 84.9% |
| ReAcTable (text-davinci-003) | 81.2% |
| with s-vote | 83.1% |
| with t-vote | 82.9% |
| with e-vote | 83.6% |
| ReAcTable (gpt3.5-turbo) | 73.1% |
| with s-vote | 72.8% |
| with t-vote | 74.4% |
| with e-vote | N.A.[1] |

***Results.*** The results of ReAcTable with various LLMs are shown in Table 10 and Table 11 for the WikiTQ and TabFact data sets, respectively. In these tables, we do not report the performance of *ReAcTable with e-vote* using *gpt3.5-turbo* since *gpt3.5-turbo* does not provide probability scores [43].

From both tables, the accuracy achieved with the additional two models (*text-davinci-003* and *gpt3.5-turbo*) is lower than when using the default Codex model (*code-davinci-002*). Specifically, with *text-davinci-003*, ReAcTable achieves a maximum accuracy of 65.0% on WikiTQ (*ReAcTable with e-vote*) and up to 52.5% on TabFact (*ReAcTable with t-vote*). Similarly, for TabFact, *ReAcTable with text-davinci-003* reaches a maximum accuracy of 83.6%, whereas *ReAcTable with gpt3.5-turbo* achieves a maximum accuracy of 74.4%. From these results, among the three models, ReAcTable performs the best when using the Codex model (*code-davinci-002*), while the accuracy is lowest when using the chat-based model (*gpt3.5-turbo*).

We also find from the results that chat-based models tend to generate answers in a more natural language form. For instance, in the case of WikiTQ, when the gold answer is "Francisco Bravo Medical Magnet High School|2007", *ReAcTable with gpt3.5-turbo* predicts "the first school to reach 800 API is Francisco Bravo Medical Magnet High School in the year 2007". While the generated answer is technically correct, it may not be compatible with our structured answer evaluator due to its non-standard format [47].

Furthermore, it is worth noting that for all ReAcTable configurations with *text-davinci-003*, execution-based voting methods (*ReAcTable with s-vote*) consistently yield the best results. This observation suggests that execution-based voting is effective in selecting semantically correct code for a model that is not primarily designed for code generation, like *text-davinci-003*.

***Takeaways.*** From the above results, ReAcTable exhibits the capability to effectively collaborate with various language models, including both chat-based models like *gpt3.5-turbo*, text-generation-based models like *text-davinci-003*, and dedicated code-generation models like *code-davinci-002*. However, the choice of model can notably impact ReAcTable's performance. Additionally, there isn't a universally optimal majority voting mechanism applicable to every model, which calls for further investigations into determining the most suitable voting mechanism for a given LLM.

## 5 LESSONS LEARNED

In this section, we summarize the lessons we learned during our exploration of LLMs in TQA tasks. We provide discussions on the design spaces that we considered and the limitations of ReAcTable.

### 5.1 Design Spaces of LLM for TQA tasks

Previous works in TQA have mainly focused on addressing the TQA task with two distinct categories: (i) approaches involving training or fine-tuning, and (ii) approaches utilizing pre-trained LLMs without training.
- **Training or fine-tuning LLMs.** Training or fine-tuning LLMs for TQA is an approach that has the potential to yield improved results. However, fine-tuning an LLM poses significant challenges, including the difficulty of acquiring training data and the high training cost.
- **Using pre-trained LLMs.** Another category pertains to addressing the TQA task using pre-trained LLMs without updating their parameters. These pre-trained LLMs, originally not tailored for the TQA task, might display relatively lower performance unless they are meticulously prompted and guided. Consequently, these approaches require specific design considerations to maximize their effectiveness.

Due to the simplicity and versatility associated with the utilization of pre-trained LLMs, we chose to explore the second category and examine how commonly used LLM enhancement frameworks, COT and ReAct, can be tailored to the TQA task. The results indicate that our ReAcTable framework can surpass many of the baseline approaches, while retaining its advantageous characteristics, including simplicity and adaptability.

### 5.2 Choosing External Code Executors

In our experiments, we employ SQL and Python as the two external code executors. We have chosen these two executors because they are commonly utilized by data scientists for tabular data manipulation tasks. In our configuration, SQL primarily handles data selection, while Python is employed for complex data reformatting operations. Furthermore, incorporating these code executors also facilitates the possibility for data scientists to easily assess the correctness of (intermediate) code. In addition, using these code executors does not introduce additional implementation complexities, unlike approaches like Binder [5], which necessitate complex re-implementation of the SQL executor.

### 5.3 Majority Voting Mechanisms

In this paper, we investigate the utilization of three majority voting methods in ReAcTable: 1) simple majority voting, 2) tree-exploration voting, and 3) execution-based voting. As elaborated in Section 3.5, simple majority voting and tree-exploration voting involve exploring multiple solution paths for the same question, whereas execution-based voting prioritizes the generation of semantically correct code.

In addition to the advantages, it is important to emphasize that the application of majority voting may potentially lead to a performance decline, particularly when the model exhibits uncertainty regarding the answer. This could be attributed to the fact that the high-temperature setup in majority voting might further accentuate the uncertainty. Furthermore, majority voting introduces additional prompting costs. Such cost can also be crucial, as prompting LLMs consumes substantial GPU resources. Therefore, one should exercise caution when considering the use of majority voting methods.

### 5.4 Limitations and Future Works

ReAcTable employs few-shot prompting methods, which necessitate manual creation of examples based on the training set. This creation pipeline can be non-trivial, and the process of tuning the prompt falls outside the scope of this paper. We identify automatic prompt tuning and the selection of few-shot examples as potential avenues for future research. Additionally, we currently focus on scenarios where the input is a single relational table. While ReAcTable has the potential to be extended for use with multiple tables, we leave this as a topic for future work. Another area for future exploration is automatic selection of the best-performing majority voting method as discussed in Section 3.5.

---
[1]Since *gpt3.5-turbo* does not provide probability scores [43], *e-vote* is not applicable on *gpt3.5-turbo*.

## 6 RELATED WORK

### 6.1 Large Language Models

In recent years, the field of natural language processing has witnessed significant advancements in large language models capable of few-shot prompting. BERT [7] introduced the concept of pre-training on vast amounts of text data followed by fine-tuning for specific tasks. GPT-2 [33] demonstrated the power of autoregressive language modeling and prompted the development of even larger models. Subsequently, GPT-3 [2] emerged as a groundbreaking model with 175 billion parameters, showcasing remarkable few-shot capabilities across a wide range of tasks. Building upon the success of GPT-3, GPT-3.5 extended the boundaries of few-shot learning [28]. The latest iteration, GPT-4 [27], further refines these capabilities, highlighting the ongoing progress in the field. In addition to these established models, more recent entrants like Code Llama [37], Claude [6] have pushed the boundaries of LLMs, underscoring the continual evolution of LLMs in addressing diverse natural language understanding and generation challenges.

### 6.2 Prompting Large Language Models

Few-shot prompting plays a pivotal role in harnessing the power of LLMs for a wide range of natural language understanding and generation tasks. By providing LLMs with a limited set of examples and a carefully crafted prompt, researchers have demonstrated their ability to generalize knowledge and perform specific tasks with minimal supervision [2]. The effectiveness of few-shot prompting is also related to prompt engineering and example selection strategies. Prompt tuning techniques, such as gradient-based optimization of prompts [21, 55], enable fine-grained control over model behavior by automatically tinning the prompt for a given task. Additionally, advanced methods for tuning the selection of few-shot examples from a large corpus [19, 30, 45], have been proposed to enhance model performance. These methods collectively enable practitioners to tailor LLMs to diverse tasks and domains, effectively leveraging the few-shot capabilities of these models to tackle real-world language understanding and generation challenges.

To further enhance the potential of LLMs in solving complex tasks, researchers have developed various prompting paradigms for LLMs. One of the leading prompting methods is the Chain-of-Thought (CoT) [46], which encourages LLMs to break down the question-answering process into multiple smaller tasks. Building upon the CoT paradigm, Tree-of-Thought [48] and Graph-of-Thought [1] have been proposed to explore multiple reasoning chains. Further extending the CoT approach, the ReAct framework [49] introduces the use of multiple external action executors, significantly enhancing the capabilities of LLMs.

### 6.3 Natural Language to Code

Natural language to code refers to the challenging task of automatically translating human-readable natural language descriptions into executable programming code. Traditionally, this problem has been tackled using rule-based approaches, which involve designing handcrafted grammar and syntactic rules to parse and interpret natural language queries [11, 14, 32].

While these rule-based approaches have shown promise, they often struggle with handling the nuances and variations of natural language. In recent years, there has been a significant shift towards model-based methods for natural language to code conversion. These methods employ machine learning models, often based on neural networks, to learn the mapping between natural language and code from large data sets [8, 9, 40].

Furthermore, the advent of large language models (LLMs) has opened up new possibilities for natural language to code tasks. Researchers have explored the use of LLMs with few-shot prompting to generate code snippets from natural language descriptions, effectively treating the model as a powerful code generator [2, 3]. To ensure the generated code is free of syntax errors, additional techniques such as controlled decoding and post-processing checks are employed [17, 18, 26, 38, 39]. These advancements highlight the potential of LLMs in automating the natural language to code conversion process, making it more robust for various applications.

The availability of high-quality natural language to code data sets, including Spider [52], CoSQL [51], SParC [53], WikiSQL [56], and BIRD [20], has been instrumental in advancing research in the domain of natural language to code conversion. These datasets provide rich and diverse examples of natural language queries paired with corresponding code snippets, covering a wide range of programming languages and database domains.

### 6.4 Table Question Answering

Table question answering (TQA), a specific variant of the general question-answering task, distinguishes itself by specializing on structured tabular data. The work in [16] presents a survey of approaches developed for this task. Recent approaches typically fall into two categories. The first category involves training or fine-tuning custom models to generate answers based on input tables. Notable methods falling within this category include Tapas [12], Tapex [23], Tacube [57], and OmniTab [15]. The second category leverages LLMs' ability to generate code for manipulating tabular data [5, 26, 50], a category to which ReAcTable also belongs. ReAcTable differentiates itself from these approaches by combining intermediate table generation along with reasoning and external tools to enhance the predictive quality of LLMs.

## 7 CONCLUSION

In this paper, we investigate how the TQA problem can be effectively addressed using foundational advances in LLMs. We introduce ReAcTable, a framework that employs LLMs to reason step-by-step and iteratively generates intermediate tables using external code executors. Our experimental results demonstrate that ReAcTable outperforms existing state-of-the-art approaches. Our findings illustrate that a simple, yet carefully adapted LLM-based framework can still surpass many state-of-the-art approaches tailored to the TQA task. We also identify several promising directions for future research, including the selection of the best majority voting method, automatic prompt tuning, and efficient TQA over multiple tables.

# REFERENCES

[1] Maciej Besta, Nils Blach, Ales Kubicek, Robert Gerstenberger, Lukas Gianinazzi, Joanna Gajda, Tomasz Lehmann, Michal Podstawski, Hubert Niewiadomski, Piotr Nyczyk, and Torsten Hoefler. 2023. Graph of Thoughts: Solving Elaborate Problems with Large Language Models. arXiv:2308.09687 [cs.CL]

[2] Tom B. Brown, Benjamin Mann, Nick Ryder, Melanie Subbiah, Jared Kaplan, Prafulla Dhariwal, Arvind Neelakantan, Pranav Shyam, Girish Sastry, Amanda Askell, Sandhini Agarwal, Ariel Herbert-Voss, Gretchen Krueger, Tom Henighan, Rewon Child, Aditya Ramesh, Daniel M. Ziegler, Jeffrey Wu, Clemens Winter, Christopher Hesse, Mark Chen, Eric Sigler, Mateusz Litwin, Scott Gray, Benjamin Chess, Jack Clark, Christopher Berner, Sam McCandlish, Alec Radford, Ilya Sutskever, and Dario Amodei. 2020. Language Models are Few-Shot Learners. arXiv:2005.14165 [cs.CL]

[3] Mark Chen, Jerry Tworek, Heewoo Jun, Qiming Yuan, Henrique Ponde de Oliveira Pinto, Jared Kaplan, Harri Edwards, Yuri Burda, Nicholas Joseph, Greg Brockman, Alex Ray, Raul Puri, Gretchen Krueger, Michael Petrov, Heidy Khlaaf, Girish Sastry, Pamela Mishkin, Brooke Chan, Scott Gray, Nick Ryder, Mikhail Pavlov, Alethea Power, Lukasz Kaiser, Mohammad Bavarian, Clemens Winter, Philippe Tillet, Felipe Petroski Such, Dave Cummings, Matthias Plappert, Fotios Chantzis, Elizabeth Barnes, Ariel Herbert-Voss, William Hebgen Guss, Alex Nichol, Alex Paino, Nikolas Tezak, Jie Tang, Igor Babuschkin, Suchir Balaji, Shantanu Jain, William Saunders, Christopher Hesse, Andrew N. Carr, Jan Leike, Josh Achiam, Vedant Misra, Evan Morikawa, Alec Radford, Matthew Knight, Miles Brundage, Mira Murati, Katie Mayer, Peter Welinder, Bob McGrew, Dario Amodei, Sam McCandlish, Ilya Sutskever, and Wojciech Zaremba. 2021. Evaluating Large Language Models Trained on Code. arXiv:2107.03374 [cs.LG]

[4] Wenhu Chen, Hongmin Wang, Jianshu Chen, Yunkai Zhang, Hong Wang, Shiyang Li, Xiyou Zhou, and William Yang Wang. 2020. TabFact: A Large-scale Dataset for Table-based Fact Verification. International Conference on Learning Representations (2020). https://openreview.net/forum?id=rkeJRhNYDH

[5] Zhoujun Cheng, Tianbao Xie, Peng Shi, Chengzu Li, Rahul Nadkarni, Yushi Hu, Caiming Xiong, Dragomir Radev, Mari Ostendorf, Luke Zettlemoyer, Noah A. Smith, and Tao Yu. 2023. Binding Language Models in Symbolic Languages. arXiv:2210.02875 [cs.CL]

[6] Claude models 2023. Claude models. Retrieved Sep 27, 2023 from https://www.anthropic.com/index/introducing-claude

[7] Jacob Devlin, Ming-Wei Chang, Kenton Lee, and Kristina Toutanova. 2018. Bert: Pre-training of deep bidirectional transformers for language understanding. arXiv preprint arXiv:1810.04805 (2018).

[8] Zhangyin Feng, Daya Guo, Duyu Tang, Nan Duan, Xiaocheng Feng, Ming Gong, Linjun Shou, Bing Qin, Ting Liu, Daxin Jiang, et al. 2020. Codebert: A pre-trained model for programming and natural languages. arXiv preprint arXiv:2002.08155 (2020).

[9] Xiaodong Gu, Hongyu Zhang, and Sunghun Kim. 2018. Deep code search. In Proceedings of the 40th International Conference on Software Engineering. 933–944.

[10] Zihui Gu, Ju Fan, Nan Tang, Preslav Nakov, Xiaoman Zhao, and Xiaoyong Du. 2022. PASTA: Table-Operations Aware Fact Verification via Sentence-Table Cloze Pre-training. arXiv:2211.02816 [cs.CL]

[11] Tihomir Gvero and Viktor Kuncak. 2015. Synthesizing Java expressions from free-form queries. In Proceedings of the 2015 acm sigplan international conference on object-oriented programming, systems, languages, and applications. 416–432.

[12] Jonathan Herzig, Pawel Krzysztof Nowak, Thomas Müller, Francesco Piccinno, and Julian Eisenschlos. 2020. TaPas: Weakly Supervised Table Parsing via Pre-training. In Proceedings of the 58th Annual Meeting of the Association for Computational Linguistics. Association for Computational Linguistics. https://doi.org/10.18653/v1/2020.acl-main.398

[13] Richard D Hipp. 2020. SQLite. https://www.sqlite.org/index.html

[14] Reid Holmes and Gail Murphy. 2005. Using structural context to recommend source code examples. Proceedings - 27th International Conference on Software Engineering, ICSE05, 117–125. https://doi.org/10.1109/ICSE.2005.1553554

[15] Zhengbao Jiang, Yi Mao, Pengcheng He, Graham Neubig, and Weizhu Chen. 2022. OmniTab: Pretraining with Natural and Synthetic Data for Few-shot Table-based Question Answering. arXiv:2207.03637 [cs.CL]

[16] Nengzheng Jin, Joanna Siebert, Dongfang Li, and Qingcai Chen. 2022. A Survey on Table Question Answering: Recent Advances. arXiv:2207.05270 [cs.CL]

[17] Rogers Jeffrey Leo John, Dylan Bacon, Junda Chen, Ushmal Ramesh, Jiatong Li, Deepan Das, Robert V. Claus, Amos Kendall, and Jignesh M. Patel. 2023. DataChat: An Intuitive and Collaborative Data Analytics Platform. In Companion of the 2023 International Conference on Management of Data, SIGMOD/PODS 2023, Seattle, WA, USA, June 18-23, 2023, Sudipto Das, Ippokratis Pandis, K. Selçuk Candan, and Sihem Amer-Yahia (Eds.). ACM, 203–215. https://doi.org/10.1145/3555041.3589678

[18] Anirudh Khatry, Joyce Cahoon, Jordan Henkel, Shaleen Deep, Venkatesh Emani, Avrilia Floratou, Sumit Gulwani, Vu Le, Mohammad Raza, Sherry Shi, Mukul Singh, and Ashish Tiwari. 2023. From Words to Code: Harnessing Data for Program Synthesis from Natural Language. arXiv:2305.01598 [cs.DB]

[19] Sawan Kumar and Partha Talukdar. 2021. Reordering examples helps during priming-based few-shot learning. arXiv preprint arXiv:2106.01751 (2021).

[20] Jinyang Li, Binyuan Hui, Ge Qu, Binhua Li, Jiaxi Yang, Bowen Li, Bailin Wang, Bowen Qin, Rongyu Cao, Ruiying Geng, et al. 2023. Can llm already serve as a database interface? a big bench for large-scale database grounded text-to-sqls. arXiv preprint arXiv:2305.03111 (2023).

[21] Xiang Lisa Li and Percy Liang. 2021. Prefix-tuning: Optimizing continuous prompts for generation. arXiv preprint arXiv:2101.00190 (2021).

[22] Chin-Yew Lin. 2004. Rouge: A package for automatic evaluation of summaries. In Text summarization branches out. 74–81.

[23] Qian Liu, Bei Chen, Jiaqi Guo, Morteza Ziyadi, Zeqi Lin, Weizhu Chen, and Jian-Guang Lou. 2022. TAPEX: Table Pre-training via Learning a Neural SQL Executor. arXiv:2107.07653 [cs.CL]

[24] Wes McKinney et al. 2010. Data structures for statistical computing in python. In Proceedings of the 9th Python in Science Conference, Vol. 445. Austin, TX, 51–56.

[25] Linyong Nan, Chiachun Hsieh, Ziming Mao, Xi Victoria Lin, Neha Verma, Rui Zhang, Wojciech Kryściński, Hailey Schoelkopf, Riley Kong, Xiangru Tang, Mutethia Mutuma, Ben Rosand, Isabel Trindade, Renusree Bandaru, Jacob Cunningham, Caiming Xiong, and Dragomir Radev. 2022. FeTaQA: Free-form Table Question Answering. Transactions of the Association for Computational Linguistics 10 (2022), 35–49.

[26] Ansong Ni, Srini Iyer, Dragomir Radev, Ves Stoyanov, Wen tau Yih, Sida I. Wang, and Xi Victoria Lin. 2023. LEVER: Learning to Verify Language-to-Code Generation with Execution. arXiv:2302.08468 [cs.LG]

[27] OpenAI. 2023. GPT-4 Technical Report. arXiv:2303.08774 [cs.CL]

[28] OpenAI models 2023. OpenAI models. Retrieved Sep 17, 2023 from https://platform.openai.com/docs/models

[29] Panupong Pasupat and Percy Liang. 2015. Compositional Semantic Parsing on Semi-Structured Tables. arXiv:1508.00305 [cs.CL]

[30] Ethan Perez, Douwe Kiela, and Kyunghyun Cho. 2021. True few-shot learning with language models. Advances in neural information processing systems 34 (2021), 11054–11070.

[31] Fotis Psallidas, Yiwen Zhu, Bojan Karlas, Jordan Henkel, Matteo Interlandi, Subru Krishnan, Brian Kroth, Venkatesh Emani, Wentao Wu, Ce Zhang, Markus Weimer, Avrilia Floratou, Carlo Curino, and Konstantinos Karanasos. 2022. Data Science Through the Looking Glass: Analysis of Millions of GitHub Notebooks and ML.NET Pipelines. SIGMOD Rec. 51, 2 (jul 2022), 30–37. https://doi.org/10.1145/3552490.3552496

[32] Chris Quirk, Raymond Mooney, and Michel Galley. 2015. Language to code: Learning semantic parsers for if-this-then-that recipes. In Proceedings of the 53rd Annual Meeting of the Association for Computational Linguistics and the 7th International Joint Conference on Natural Language Processing (Volume 1: Long Papers). 878–888.

[33] Alec Radford, Jeffrey Wu, Rewon Child, David Luan, Dario Amodei, Ilya Sutskever, et al. 2019. Language models are unsupervised multitask learners. OpenAI blog 1, 8 (2019), 9.

[34] Colin Raffel, Noam Shazeer, Adam Roberts, Katherine Lee, Sharan Narang, Michael Matena, Yanqi Zhou, Wei Li, and Peter J. Liu. 2020. Exploring the Limits of Transfer Learning with a Unified Text-to-Text Transformer. Journal of Machine Learning Research 21, 140 (2020), 1–67. http://jmlr.org/papers/v21/20-074.html

[35] Colin Raffel, Noam Shazeer, Adam Roberts, Katherine Lee, Sharan Narang, Michael Matena, Yanqi Zhou, Wei Li, and Peter J. Liu. 2023. Exploring the Limits of Transfer Learning with a Unified Text-to-Text Transformer. arXiv:1910.10683 [cs.LG]

[36] Joshua Robinson, Christopher Michael Rytting, and David Wingate. 2022. Leveraging large language models for multiple choice question answering. arXiv preprint arXiv:2210.12353 (2022).

[37] Baptiste Rozière, Jonas Gehring, Fabian Gloeckle, Sten Sootla, Itai Gat, Xiaoqing Ellen Tan, Yossi Adi, Jingyu Liu, Tal Remez, Jérémy Rapin, Artyom Kozhevnikov, Ivan Evtimov, Joanna Bitton, Manish Bhatt, Cristian Canton Ferrer, Aaron Grattafiori, Wenhan Xiong, Alexandre Défossez, Jade Copet, Faisal Azhar, Hugo Touvron, Louis Martin, Nicolas Usunier, Thomas Scialom, and Gabriel Synnaeve. 2023. Code Llama: Open Foundation Models for Code. arXiv:2308.12950 [cs.CL]

[38] Torsten Scholak, Nathan Schucher, and Dzmitry Bahdanau. 2021. PICARD: Parsing Incrementally for Constrained Auto-Regressive Decoding from Language Models. In Proceedings of the 2021 Conference on Empirical Methods in Natural Language Processing. Association for Computational Linguistics, 9895–9901. https://aclanthology.org/2021.emnlp-main.779

[39] Freda Shi, Daniel Fried, Marjan Ghazvininejad, Luke Zettlemoyer, and Sida I Wang. 2022. Natural language to code translation with execution. arXiv preprint arXiv:2204.11454 (2022).

[40] Ilya Sutskever, Oriol Vinyals, and Quoc V Le. 2014. Sequence to sequence learning with neural networks. Advances in neural information processing systems 27 (2014).

[41] Temperature setup of GPT models 2023. Temperature setup of GPT models. Retrieved Sep 28, 2023 from https://platform.openai.com/docs/api-reference/completions

[42] The Codex model of OpenAI 2023. *The Codex model of OpenAI*. Retrieved Sep 28, 2023 from https://openai.com/blog/openai-codex
[43] Usage of GPT-3.5-Turbo and GPT-4 Models 2023. *Usage of GPT-3.5-Turbo and GPT-4 Models*. Retrieved Sep 17, 2023 from https://learn.microsoft.com/en-us/azure/ai-services/openai/how-to/chatgpt?pivots=programming-language-chat-completions
[44] Ashish Vaswani, Noam Shazeer, Niki Parmar, Jakob Uszkoreit, Llion Jones, Aidan N. Gomez, Lukasz Kaiser, and Illia Polosukhin. 2023. Attention Is All You Need. arXiv:1706.03762 [cs.CL]
[45] Yikai Wang, Chengming Xu, Chen Liu, Li Zhang, and Yanwei Fu. 2020. Instance credibility inference for few-shot learning. In *Proceedings of the IEEE/CVF conference on computer vision and pattern recognition*. 12836–12845.
[46] Jason Wei, Xuezhi Wang, Dale Schuurmans, Maarten Bosma, Brian Ichter, Fei Xia, Ed Chi, Quoc Le, and Denny Zhou. 2023. Chain-of-Thought Prompting Elicits Reasoning in Large Language Models. arXiv:2201.11903 [cs.CL]
[47] WikiTableQuestions data set 2023. *WikiTableQuestions data set*. Retrieved Sep 28, 2023 from https://github.com/ppasupat/WikiTableQuestions
[48] Shunyu Yao, Dian Yu, Jeffrey Zhao, Izhak Shafran, Thomas L. Griffiths, Yuan Cao, and Karthik Narasimhan. 2023. Tree of Thoughts: Deliberate Problem Solving with Large Language Models. arXiv:2305.10601 [cs.CL]
[49] Shunyu Yao, Jeffrey Zhao, Dian Yu, Nan Du, Izhak Shafran, Karthik Narasimhan, and Yuan Cao. 2023. ReAct: Synergizing Reasoning and Acting in Language Models. arXiv:2210.03629 [cs.CL]
[50] Yunhu Ye, Binyuan Hui, Min Yang, Binhua Li, Fei Huang, and Yongbin Li. 2023. Large Language Models are Versatile Decomposers: Decompose Evidence and Questions for Table-based Reasoning. arXiv:2301.13808 [cs.CL]
[51] Tao Yu, Rui Zhang, He Yang Er, Suyi Li, Eric Xue, Bo Pang, Xi Victoria Lin, Yi Chern Tan, Tianze Shi, Zihan Li, et al. 2019. Cosql: A conversational text-to-sql challenge towards cross-domain natural language interfaces to databases. *arXiv preprint arXiv:1909.05378* (2019).
[52] Tao Yu, Rui Zhang, Kai Yang, Michihiro Yasunaga, Dongxu Wang, Zifan Li, James Ma, Irene Li, Qingning Yao, Shanelle Roman, et al. 2018. Spider: A large-scale human-labeled dataset for complex and cross-domain semantic parsing and text-to-sql task. *arXiv preprint arXiv:1809.08887* (2018).
[53] Tao Yu, Rui Zhang, Michihiro Yasunaga, Yi Chern Tan, Xi Victoria Lin, Suyi Li, Heyang Er, Irene Li, Bo Pang, Tao Chen, et al. 2019. Sparc: Cross-domain semantic parsing in context. *arXiv preprint arXiv:1906.02285* (2019).
[54] Minjia Zhang, Conglong Li, Xiaoxia Wu, Zhewei Yao, and Yuxiong He. 2022. SaMoE: Parameter Efficient MoE Language Models via Self-Adaptive Expert Combination. (2022).
[55] Ningyu Zhang, Luoqiu Li, Xiang Chen, Shumin Deng, Zhen Bi, Chuanqi Tan, Fei Huang, and Huajun Chen. 2021. Differentiable prompt makes pre-trained language models better few-shot learners. *arXiv preprint arXiv:2108.13161* (2021).
[56] Victor Zhong, Caiming Xiong, and Richard Socher. 2017. Seq2sql: Generating structured queries from natural language using reinforcement learning. *arXiv preprint arXiv:1709.00103* (2017).
[57] Fan Zhou, Mengkang Hu, Haoyu Dong, Zhoujun Cheng, Shi Han, and Dongmei Zhang. 2022. TaCube: Pre-computing Data Cubes for Answering Numerical-Reasoning Questions over Tabular Data. arXiv:2205.12682 [cs.IR]
[58] Wenhao Zhu, Hongyi Liu, Qingxiu Dong, Jingjing Xu, Shujian Huang, Lingpeng Kong, Jiajun Chen, and Lei Li. 2023. Multilingual Machine Translation with Large Language Models: Empirical Results and Analysis. arXiv:2304.04675 [cs.CL]